# ADS as Information Management Service in an M-Learning Environment

Matthias R. Brust[1], Daniel Görgen[2], Christian Hutter[1], and Steffen Rothkugel[1]

[1] Université du Luxembourg ⋆ ⋆ ⋆
Faculté des Sciences, de la Technologie et de la Communication
Campus Kirchberg, 6, rue Coudenhove-Kalergi
L-1359 Luxembourg-Kirchberg, Luxembourg
{Matthias.Brust, Christian.Hutter, Steffen.Rothkugel}@univ.lu
[2] University of Trier †
System Software and Distributed Systems
Universitätsring
D-54286 Trier, Germany
Goergen@syssoft.uni-trier.de

**Abstract.** Leveraging the potential power of even small handheld devices able to communicate wirelessly requires dedicated support. In particular, collaborative applications need sophisticated assistance in terms of querying and exchanging different kinds of data. Using a concrete example from the domain of mobile learning, the general need for information dissemination is motivated. Subsequently, and driven by infrastructural conditions, realization strategies of an appropriate middleware service based upon an information market model are discussed.

## 1 Introduction

Nowadays, a large variety of more and more powerful mobile devices like smartphones, personal digital assistants, and laptops is available. The increase of computational power and memory space allows the adaptation of applications from different domains into the realm of mobility. As most of those devices are shipped with wireless networking adapters, it is reasonable to use those for communication purposes. It would also be possible to use technologies like GSM or UMTS, but due to the costs induced they should be employed only where appropriate. Also it is not reasonable to use a third party like a phone company to communicate with someone being close or even in the same room. By using their wireless connection adapters the devices can communicate directly and are independent from stationary nodes.

Low bandwidth, unreliability of wireless links, small transmission ranges, and unpredictable network topology changes, however, pose technically challenging problems.

⋆ ⋆ ⋆ This research is funded by the Luxembourg Ministre de la Culture, de l'Enseignement Suprieur et de la Recherche.
† This work is funded in part by DFG, Schwerpunktprogramm SPP1140 "Basissoftware für selbstorganisierende Infrastrukturen für vernetzte mobile Systeme"

In ad-hoc networks, resources and services need to be spread across a bigger area than just the transmission range of a single device. Hence, devices are required to communicate with one another by routing information through intermediate nodes, resulting in so-called multi-hop ad-hoc networks. Besides popular gaming applications which represent a growing market using platforms like Nokia's N-Gage, such network infrastructures can also be exploited by other application domains. One example is mobile learning, or M-learning, a growing and challenging area. This paper introduces CARLA, an M-Learning application scenario, and discusses its use on top of ADS, a generic information management service for multi-hop ad-hoc networks.

The structure of this paper is as follows. The following section introduces CARLA and motivates the usefulness of ADS in this context. In the subsequent section, basic concepts of ADS are discussed. Section four compares ADS to similar middleware services. Finally, a summary is given in section five, mentioning aspects of ongoing and future work.

## 2   CARLA, an M-Learning Application Scenario

Using mobile devices like PDAs becomes more commonplace. It is certainly reasonable to use them for improving the learning process [1]. CARLA is a distributed learning application designed for mobile devices equipped with wireless communication adapters. Students can use the system during and after lectures, being able to join forces by sharing their material, and can help each other in a cooperative way, e.g. to prepare for exams.

CARLA can primarily be used by students to manage teaching material like lecture slides and articles. The teaching staff, i.e. professors, teachers, and tutors, uses the system to distribute such material among their students. Initially, all students basically have access to the same material. Possible locations of initial releases could be lecture halls and staff offices. Some students, however, might only have received subsets of the material released, for instance because of not having attended to a lecture. CARLA explicitly enables them to capture missing parts from their fellows later.

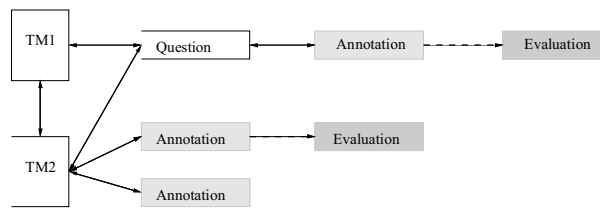

**Fig. 1.** Personalized View of Different Learning Material

During a lecture, a student's device can support him by showing the appropriate slides. Depending on the display capacities of a device this might be the real slide or, paying respect to the complexity of the slide, only some abstract representation. In any case, students are encouraged to add *annotations* to a slide.

While recapitulating the material later on, students can use *multiple-choice questions* to get a deeper understanding of the topics covered. The questions can be provided by the staff or by other students. Moreover, students might discover additional relationships between some sections of the teaching material, their annotations, and the questions, adding them to their material. In CARLA, such relationships are called *links*.

CARLA allows students to have personalized sets of their teaching material, as illustrated in in Fig. 1, including annotations, questions and links. During runtime, the subset of available material can be augmented by meeting other students.

To prevent misleading or false additions to the teaching material from being distributed, each student is encouraged to evaluate such material received by others. This allows CARLA to detect and remove fakes, trying to improve the usefulness of the data.

To boost the students' motivation, the teaching staff can use CARLA to create a quiz where the students act as players getting points for answering questions correctly. Similar to well-known TV game shows, players can use jokers for receiving hints. Three different kinds of jokers are supported: *link jokers*, *annotation jokers*, and *statistics jokers*. Link jokers allow exposing links pointing to additional material. Annotation jokers reveal annotations for a question, and statistics jokers can be used to show statistics indicating how other players answered this particular question, based on the available information. At a predefined time, e.g. at the last lecture of a course, the player with the highest rank wins.

Besides improving the learning potential in general, CARLA also fosters the work of the teaching staff. Influencing the learning process is possible at any time by adding additional links and/or annotations. Furthermore, CARLA allows redesigning the initial teaching material based on the students' feedback given by links, questions and annotations, thereby increasing its scope and usefulness.

Managing the teaching material together with the data added by the students is a challenging task. All devices participating, both of staff and students, form multiple ad-hoc network partitions over time. Particularly, there is neither a central network backbone, nor persistent and reliable communication channels. Students cannot simply retrieve additional or publish new material just by establishing a connection to a central server. Nevertheless, the teaching material and the material added by the students like questions, annotations and evaluations needs to be shared among all CARLA users. An analysis of the application scenario results in the general requirements of distributed applications for mobile, multi-hop ad-hoc networks.

## 3   Information Dissemination and Sharing in Mobile Ad-hoc Networks

Similar to distributed applications for traditional wired networks, the ultimate objective of applications running in mobile multi-hop ad-hoc networking environments is to have all relevant and useful data available at all times. Obviously, this goal cannot, and in fact need not, be reached to a full extent. However, the approach is to optimize the process of gathering data as far as possible.

Due to the characteristics of the environment envisaged, traditional strategies involving backbone network connections respectively links to arbitrary devices cannot be applied. For instance, it is not feasible for an application missing some data simply to use push or pull strategies to fetch the data from central servers. The more or less unpredictable and frequently changing topology of mobile ad-hoc networks requires special support by the underlying middleware.

The task of this middleware is to try to gather as much relevant information as possible to make it available locally for all applications running on that device. Applications thus will be enable to receive information fast and synchronously by querying the local device. Additionally, in still adhering to the synchronous paradigm, the data might be augmented by additionally querying the immediate neighborhood. In the M-Learning scenario introduced before, CARLA stores parts of the teaching material on the local device and can query nearby devices for additional or missing data, resulting in a local subset of the globally available information.

There might remain, however, a gap between the data gathered locally and the global set of information. Hence, the middleware should also support additional ways to query for this kind of relevant data stored on devices not reachable directly. Queries in these cases obey an asynchronous nature. After being initiated, queries can be seen as active entities, returning results sporadically and potentially multiple times. In CARLA, a query for missing teaching material, additional questions and more might be initiated when a student arrives at the university, continuing to gather data over a given period of time, for instance until returning back home.

There are different strategies to realize such a middleware. The idea discussed below is based upon so-called *information markets*. In the subsequent section, basic concepts of ADS, a middleware service supporting information dissemination, will be introduced.

## 4 ADS—A Distributed Information Service for Mobile Ad-hoc Networks

ADS is an approach for a fully distributed information service for multi-hop ad hoc networks, based upon an information market model. It allows the sharing of potentially replicated information among applications running on the nodes of such networks. ADS is designed as a middleware service running on all devices participating. Applications using ADS in turn are supposed to act in a cooperative way. The data those applications operate on is initially generated on a particular device. This information might be useful for the local device only, or might be shared with other instances of the application, respectively with other users. In the latter case, the need for disseminating the information in a controlled way arises. Before focusing on some basic ADS concepts, the topology of the underlying network infrastructure is presented in the following paragraph.

### 4.1 Ad-hoc Network Infrastructure

Ad-hoc networks formed by a large number of mobile devices tend to have recurrent and long-term permanent patterns. It is possible to identify network regions regularly having high device density, so-called *hotspots*. These are formed by a potentially varying set of devices being located at specific geographic areas. The occurrence of these regions is a general observation.

For instance with respect to the M-Learning scenario, ad-hoc networks are formed mostly at universities and schools. On the one hand, there are the application-independent hotspots such as cafeterias, libraries, meeting areas and building entries which are typically populated by students of different faculties. In such hotspots, the fluctuation is comparatively high, but the average device density can be assumed to stay above a critical threshold.

On the other hand, more tentative hotspots can be identified, like lecture halls and rooms where students following the same studies work together or do their homework. The latter students typically use a common set of applications, sharing similar data. In these kinds of hotspots, almost stable ad-hoc networks are established typically for short durations, but in a recurrent manner.

In general, hotspots do not have permanent network connections with each other. They should be considered as separate network partitions. It is common, however, that data can be exchanged across hotspots, and finally between devices and a hotspot. This is due to devices located in between hotspots as well as due to mobility and people walking along corridors, stairways or footways, being able to carry information from one place to another.

To detect recurrent topological patterns as well as driving the process of information exchange, it is necessary that devices are able to determine their position, respectively the room they are currently located in. This can be achieved by outdoor positioning techniques such as GPS in conjunction with indoor approaches based on RF or IR tags. Using position determination, it becomes possible to dynamically and continuously learn about these geographic topologies and keep them up-to-date.

The infrastructure as envisaged is optionally augmented by so-called *support nodes*. These are comparatively small stationary devices equipped with wireless communication adapters. They are not intended to serve as compute nodes, but mainly for providing storage capacities. Support nodes are designed to be self-organizing and not to rely on a network backbone. One specific duty of support nodes is to assist in infrastructure management, for instance with respect to geographic information.

### 4.2 Synchronous Information Retrieval

The purpose of *synchronous queries* is to provide information which is available immediately. The data returned by synchronous queries is composed of the data available locally, and might be augmented optionally by querying the immediate neighborhood.

The data stored locally is a combination of input provided by applications or services running on the device itself together with information gathered before from other devices. Using a profile created and maintained by the local applications, each device transparently starts updating its information when getting into communication range of other devices. This en-passant communication is already reasonable during short periods of interaction, as it is not mandatory to exchange all relevant information between two devices. For instance in the M-Learning scenario, even a subset of the newly available annotations, questions, and links is useful for students. Additional data might still be gathered when meeting yet more fellows.

Optionally, the data available locally might be augmented by query results from nearby devices. This information gathering process is employing a timeout mechanism to cope with the properties of ad-hoc networks properly, and relies on short range communication within a geographic cluster where topology based routing strategies can be applied [2].

### 4.3 Asynchronous Information Retrieval

The amount of information available directly through synchronous communication is potentially limited. Therefore, more powerful sources of information are required. A possible approach to receive non-local information is the usage of information markets in combination with *asynchronous smart remote queries*. For instance in CARLA, students planning not to attend to a lecture can get the new material from an information market later. Before explaining further details about asynchronous smart remote queries, however, the concept of information markets needs to be introduced.

### 4.4 Information Markets

Due to the very dynamic structure of the network and the potentially high number of interacting devices, it is neither sensible nor possible to directly query any device. Furthermore, in pure ad-hoc networks there is no notion of central servers. Hence, strategies for collecting, exchanging, and gathering information are required.

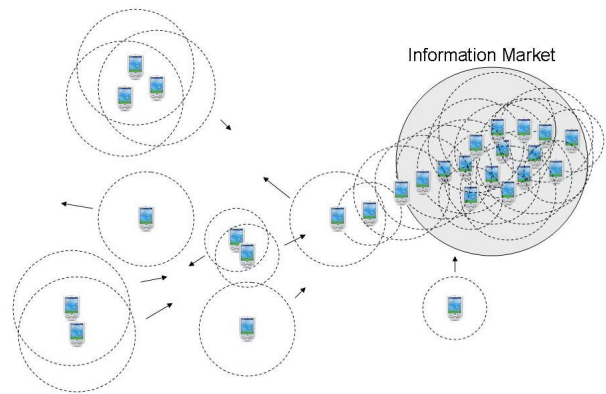

**Fig. 2.** Example Mobile Ad-hoc Network Scenario Including an Information Market

The approach proposed in this paper for tackling this problem is to exploit existing characteristics of real-life ad-hoc network infrastructure, particularly hotspots. Selected hotspots can be employed as so-called information markets, concentrating and managing large amounts of information in an adequate limited region, making it accessible to interested applications, as illustrated in Fig. 2.

One challenge inherent to this approach is the localization of the markets. In this respect, it is possible to benefit from the infrastructure as introduced before. In particular, support nodes can provide the location of information markets to interested devices. During runtime, a device might nevertheless discover additional, respectively newly created markets, particularly their location and available information categories. Descriptions of information markets are further disseminated by propagating them to other devices and forwarding them to other markets as well as support nodes.

Publishing new information on information markets is facilitated by using geographic routing strategies. Applications or services select an appropriate information market by applying context-aware strategies, e.g. by aspects of best fitting in terms of information categories already managed by a market as well as by distance of targeted applications. Published information is available afterwards by sending asynchronous smart remote queries to a market.

With respect to information markets, support nodes can provide added-value in terms of preventing information loss in case of the device density dropping below a critical threshold. Support nodes, however, cannot fully replace the information markets. These are composed of a set of potentially varying devices, entering and leaving the market area. Only on top of multiple devices replication strategies can be applied to cope with hardware and software failures and to increase fault tolerance. Furthermore, distribution fosters scalability because of sharing load. This applies both to storage capacities as well as to computing load.

### 4.5 Asynchronous Smart Remote Queries

Aside from retrieving information from the immediate neighborhood, it shall be possible to consult information markets as well. Queries launched will be sent to the information market by using geographic routing strategies [3], starting to collect results there. Queries might stay active for a given time, sending back results to the initiator in chunks.

Queries as well as responses can contain meta-data. Mandatory meta-data of queries include information about the query initiator together with some contextual information about his planned movements - e.g. taken from his calendar - that is used to determine time and location of where to send results to. Optional meta-data of queries might for instance indicate an expected number of results. Knowledge about other information markets can be propagated and collected by piggybacking it in the meta-data of responses.

## 5 Related Work

As illustrated in the M-learning scenario described, the purpose of ADS is to manage a wide variety of application-dependent information classified in different categories. Simple strategies like flooding the network by disseminating data using adaptive protocols like introduced in [4] are not reasonable in general. A different approach is used in NOM [5] which forwards queries and creates responses from every node in the network. In both strategies, broadcast storms are likely to occur. SIDE Surfer [6] in turn only allows the automatic exchange between directly connected devices based on user profiles, giving the applications access to a limited set of information only. In contrast to this, ADS allows applications to use the synchronous local queries to access such information and additionally provides asynchronous smart remote queries for retrieving data from information pools currently available on information markets. As shown in [7], the communication using the market places model is a promising approach. Both TOTA ("Tuples on the air" [8]) which aims at supporting adaptive context aware activities and MeshMdl [9] use the tuple paradigm in mobile ad hoc networks. But they miss the concept of controlled information replication, thus increasing the probability of loosing important information. Systems like CachePath, CacheData or HybridCache [10] use ad-hoc routing to send requests to servers and provide caching mechanisms to improve the average access time. But the system is designed to get data from a well-known source. Replication is used only to speed up the query instead of preventing from information losses. There are no generic methods for information access. Thus applications need to know which server is providing the information they need. As shown in more detail in [11], ADS provides generic methods to access information independently from the location of their original source and provides replication strategies to improve the data availability.

## 6  Conclusion and Future Work

ADS is a middleware service designed for use in mobile multi-hop ad-hoc environments. It enables applications to directly share information as well as receive information from the devices in the neighborhood, together with providing access to information markets. The main idea of introducing information markets is to be able to identify well-known places where different kinds of information from multiple applications can be pooled and exchanged. Within information markets, sophisticated algorithms can be applied e.g. in terms of load balancing and fault tolerance by replication. The interaction with information markets is facilitated by asynchronous smart remote queries, which are long-term queries. After being launched, these queries travel to information markets, gathering information there to be sent back to their initiator. This makes the information market model well-suited for data that remains stable for a certain time.

With respect to CARLA, the M-learning application scenario introduced, the ADS concept is able to share annotations, questions, and links available directly in the current neighborhood by using synchronous information retrieval queries. However, it is often necessary to additionally receive non-local information, e.g. when students do not attend to a lecture, but still wishing to get the newly provided teaching material. For this, asynchronous smart remote queries that employ information markets are used.

The environment envisaged is neither stable nor fully predictable. The system might for instance suffer from network partitioning, resulting e.g. in queries reaching information markets either late or possibly not at all. Another problem occurs if the number of participating devices drops under a certain threshold. Then, even with the help of the support nodes, proper dissemination and replication of information within information markets cannot be guaranteed anymore. Strong guarantees cannot be given in such kind of environments anyway. It is only possible to minimize the impact of different kinds of failures and shortcomings. In the model proposed, this is done through the use of support nodes and information markets together with their management strategies.

In the future, several additional aspects need to be evaluated. For instance, the management of data within information markets is an area for future research. It might also be interesting to evaluate if an exchange of information across several information markets would improve the overall system effectiveness. Additionally, it is currently assumed that information is stored atomically. In order to leverage both fault tolerance and load balancing, it could be interesting to split one piece of information into multiple chunks obeying a certain level of redundancy, and to distribute those on several devices. Aside from running CARLA in a testbed, the system should be evaluated in a realistic environment, namely letting students use it for a class. Finally, several application domains exist that obey different interaction paradigms. Examples include cases where data becomes out-of-date rather quickly, or applications that are comparatively tightly coupled. Hence, other concepts respectively communication patterns aside from the information market model for sharing and exchanging information need to be developed.